\documentclass{ws-procs9x6-cpt16}
\begin{document}

\newcommand{\refeq}[1]{(\ref{#1})}
\def\etal {{\it et al.}}


\def\al{\alpha}
\def\be{\beta}
\def\ga{\gamma}
\def\de{\delta}
\def\ep{\epsilon}
\def\ve{\varepsilon}
\def\ze{\zeta}
\def\et{\eta}
\def\th{\theta}
\def\vt{\vartheta}
\def\io{\iota}
\def\ka{\kappa}
\def\la{\lambda}
\def\vpi{\varpi}
\def\rh{\rho}
\def\vr{\varrho}
\def\si{\sigma}
\def\vs{\varsigma}
\def\ta{\tau}
\def\up{\upsilon}
\def\ph{\phi}
\def\vp{\varphi}
\def\ch{\chi}
\def\ps{\psi}
\def\om{\omega}
\def\Ga{\Gamma}
\def\De{\Delta}
\def\Th{\Theta}
\def\La{\Lambda}
\def\Si{\Sigma}
\def\Up{\Upsilon}
\def\Ph{\Phi}
\def\Ps{\Psi}
\def\Om{\Omega}
\def\cA{{\cal A}}
\def\cB{{\cal B}}
\def\cC{{\cal C}}
\def\cE{{\cal E}}
\def\cl{{\cal L}}
\def\cL{{\cal L}}
\def\cM{{\cal M}}
\def\cO{{\cal O}}
\def\cP{{\cal P}}
\def\cR{{\cal R}}
\def\cV{{\cal V}}
\def\mn{{\mu\nu}}

\def\fr#1#2{{{#1} \over {#2}}}
\def\half{{\textstyle{1\over 2}}}
\def\frac#1#2{{\textstyle{{#1}\over {#2}}}}

\def\vev#1{\langle {#1}\rangle}
\def\expect#1{\langle{#1}\rangle}
\def\abs#1{\left|{#1}\right|}

\def\lsim{\mathrel{\rlap{\lower4pt\hbox{\hskip1pt$\sim$}}
    \raise1pt\hbox{$<$}}}
\def\gsim{\mathrel{\rlap{\lower4pt\hbox{\hskip1pt$\sim$}}
    \raise1pt\hbox{$>$}}}
\def\prt{\partial}
\def\pt#1{\phantom{#1}}
\def\ol#1{\overline{#1}}

\def\etal{{\it et al.}}

\def\Re{\hbox{Re}\,}
\def\Im{\hbox{Im}\,}

\def\lrpartial{\raise 1pt\hbox{$\stackrel\leftrightarrow\partial$}}
\def\lrprt{\stackrel{\leftrightarrow}{\partial}}
\def\lrprtmu{\stackrel{\leftrightarrow}{\partial_\mu}}
\def\lrprtnu{\stackrel{\leftrightarrow}{\partial^\nu}}
\def\lrDmu{\stackrel{\leftrightarrow}{D_\mu}}
\def\lrDnu{\stackrel{\leftrightarrow}{D^\nu}}
\def\lrvec#1{\stackrel{\leftrightarrow}{#1} }

\def\mi{m_i}
\def\mib{\ol m_i}
\def\gw{g}
\def\gs{g_{\rm s}}

\title{Lorentz and CPT Violation in Heavy Quark Physics}

\author{M.S.\ Berger}

\address{Physics Department, Indiana University, Bloomington, IN 47405, USA}

\begin{abstract}
Violations of the Lorentz and CPT symmetries can appear as observable effects in the direct 
production of top quarks and their subsequent decays. Earlier results for the $q\bar{q}$ production process for 
$t\bar{t}$ pairs have 
been extended to include the gluon fusion process which dominates at the LHC.
In addition results are obtained for testing CPT symmetry through single top quark production.
\end{abstract}

\bodymatter

\section{Introduction}
Fundamental tests of symmetry principles form a significant element of the foundation of modern physics. 
The experimental evidence for
the violation of some of these symmetries has often led to important progress in particle physics. The Lorentz and CPT 
symmetries are among the most fundamental and the best verified. At present no evidence exists
for any violations. Nevertheless it is useful to extend the experimental tests as much as possible to new unexplored areas.  

The Standard-Model Extension (SME) is one method for parameterizing the possible sources of Lorentz- and CPT-violation
involving the Standard-Model fields.\cite{Colladay:1996iz,Colladay:1998fq} It provides a systematic approach 
for listing all possible forms of Lorentz violation 
which can occur in local field theory consistent with the required gauge symmetries. 
The Lorentz and CPT-violating effects are 
encoded in the form of coefficients, and the Lorentz-violating terms are observer scalar densities formed by 
contracting the Lorentz-violating operators with associated coefficients. Theoretical calculations can convert these 
coefficients into predicted effects in physical experiments; in particular, these coefficients can be included in  new Feynman
rules which allow one to consider their effects in scattering experiments at high energy colliders. As an inventory of the 
possible Lorentz-violating effects, the SME coefficients are independent and unrelated to each other. The coefficients can 
be related to each other if theoretical ideas exist for their origin; see, e.g., Ref.\ \refcite{Berger:2012my}.
  
Collider physics experiments offer the possibility of observing the effects of Lorentz and CPT violation in the observed 
angular and momentum distributions of the produced particles. The most common predicted effect is that the experimental
deviations from the standard results will manifest themselves as sidereal signals since the preferred directions and frames 
associated with the Lorentz violation is assumed to be fixed in space as the Earth rotates. Other effects might lead to 
modifications in the overall cross sections or other time-independent modification to the distributions. Since particles are 
produced in collider experiments in a large range of momenta and orientations, one has in principle access to many 
observer frames in which the effects of Lorentz violation might manifest themselves. On the other hand one does not 
expect the bounds obtained in collider experiments to compete with other experiments that can take advantage of an
interferometric nature such as neutral-meson oscillations.\cite{meson:tp} The top quark as well as lighter quarks contribute to these 
processes in loops in the Standard Model. Such experiments might offer the best place to bound Lorentz-violation coefficients in the quark sector of the Standard Model. 

The top quark is the heaviest quark in the Standard Model with a Yukawa coupling of order one. The other quarks have
masses much less than the electroweak scale. Thus the top quark has sometimes been singled out as special in model 
building. While there is no obvious connection between the electroweak symmetry breaking and a possible breaking 
of the Lorentz and CPT symmetries, it may nevertheless prove advantageous to think of the third generation as somewhat
more interesting a target for experimental tests. In any event the lack of hadronization of the top quark gives a unique
environment (compared to the other quarks) to interpret any possible signals especially at the LHC where the largest
samples of top events are for the first time possible. The D0 Collaboration performed the first test of Lorentz violation
in $t\bar{t}$ production using Tevatron data.\cite{Abazov:2012iu}
In this proceedings contribution,
we report on the recent progress\cite{Berger:2015yha}
in extending the $t\bar{t}$ production process to include the 
gluon fusion mechanism which provides the bulk of the quark pairs at the LHC. In addition the first calculations of the 
expected modifications to single top quark and single top antiquark have been obtained for the first time. These latter 
processes are sensitive to CPT violation as well as Lorentz violation.

\section{SME coefficients}
The coefficients involving CPT-even Lorentz violation that can be bounded in top quark experiments are contained 
in the SME lagrangian,
\begin{eqnarray}
\cl^{\rm CPT+}_{t,b} &=& 
\half i (c_L)_{\mu\nu} \ol t_L \ga^\mu \lrprtnu t_L 
+ \half i (c_R)_{\mu\nu} \ol t_R \ga^\mu \lrprtnu t_R 
\nonumber \\ &&
+ \half i (c_L)_{\mu\nu} \ol b_L \ga^\mu \lrprtnu b_L
+ (\fr {\gw V_{tb}} {\sqrt{2}} (c_L)_{\mu\nu} W^{-\nu} 
\ol b_L \ga^\mu t_L + {\rm h.c.})
\nonumber \\ &&
+ \half \gs ((c_L)_\mn +(c_R)_\mn)(\ol t \ga^\mu G^\nu t  + \ol b \ga^\mu G^\nu b ) .
\label{lorviol}
\end{eqnarray}
Similarly,
the CPT-odd terms 
can be written 
\begin{eqnarray}
\cl^{\rm CPT-}_{t,b} 
&=& 
- (a_L)_\mu \ol t_L \ga^\mu t_L 
- (a_R)_\mu \ol t_R \ga^\mu t_R 
- (a_L)_\mu \ol b_L \ga^\mu b_L .
\label{cptviol}
\end{eqnarray}
These coefficients appear in squared matrix elements contracted with the physical momenta of observable particles.

\section{Calculation}
The calculation of the matrix elements for both top-antitop production as well as single top quark/antiquark production
are described in detail in a recent publication.\cite{Berger:2015yha} Here we just highlight some of the important details 
(see also Refs.\ \refcite{Berger:2010mga,Berger:2013tla,Liu:tp}).

The SME contains coefficients which contribute to new Feynman diagrams. These include insertions on propagators as well as vertex insertions for some coefficients. There are some subtle issues regarding the insertions on external legs and the identification of the asymptotic states which have 
been discussed in the literature; these are largely avoided in the present calculation as the Lorentz violation is assumed
to be confined to the third generation which is not among the directly observed particles in an actual experiment.
There is a need to perform modified spin sums for the CPT-violating coefficient considered in  Ref.\ \refcite{Berger:2015yha}; details can be found in the appendix of that paper.

At the lowest order the process $gg\to t\bar{t}$ is a textbook problem these days. The only subtlety is the removal of 
the contribution of the unphysical longitudinal modes of the gluons which can be done by using explicit polarization vectors
or using ghosts to eliminate the unwanted contributions in the squared matrix elements. These well-known techniques 
can be used in a straightforward way when including the possible Lorentz-violating effects.

The narrow width approximation is used to factorize the physics into a production process and decay processes. This is 
known to work well for top-quark physics and one can reintroduce spin-correlation effects if one desires.

Field redefinitions can be used to show that only the symmetric parts of the coefficient $c_{\mu \nu}$ are physical. This 
provides a nontrivial check of the calculation; the contributions from the antisymmetric parts of the coefficient only cancel in 
general when all vertex and propagator insertions are included. 
 
\section{Summary}
Results for the squared-matrix elements appropriate for testing Lorentz violation in top quarks produced at the LHC have
been obtained. These can be used to improve the bounds on Lorentz violation as well as extend the results to include 
CPT violation.

\section*{Acknowledgments}
This work is supported in part
by the Department of Energy under grant number {DE}-SC0010120
and by the Indiana University Center for Spacetime Symmetries (IUCSS).

\end{document}